\documentclass[final]{siamltex}
\usepackage{amsmath,amssymb,graphicx,comment}
\newtheorem{remark}[theorem]{Remark}
\newcommand{\R}{{\mathbb R}}

\title{On the blowing up of solutions to  quantum hydrodynamic models on  bounded
domains
\thanks{Irene M. Gamba is supported by NSF-DMS0507038. Maria Pia Gualdani acknowledges partial support from the
 Deutsche Forschungsgemeinschaft, grants JU359/5.
Ping Zhang is partially supported by the
 NSF of China under Grant 10525101 and
10421101, and the innovation grant from the Chinese Academy of
Sciences. Part of the work was done when Ping Zhang visited the
department of Mathematics of Texas University at Austin, the
author would like to thank the hospitality of the department.
Support from the Institute for Computational
Engineering and Sciences at the University of Texas at Austin is
also gratefully acknowledged.}}

\author{Irene M. Gamba\thanks{Department of mathematics,
University of Texas at Austin, TX, 78712, USA; e-mail: {\tt
gamba@math.utexas.edu.}} \and Maria Pia Gualdani\thanks{Department of mathematics, University of Texas at Austin, Texas 78712, USA; e-mail: {\tt gualdani@math.utexas.edu.}} \and
Ping Zhang\thanks{Academy of Mathematics and System Sciences, The
Chinese Academy of Sciences, Beijing 100080, China; e-mail: {\tt
zp@amss.ac.cn.}}}

\begin{document}

\maketitle


\begin{abstract}
The blow-up in finite time for the solutions to the initial-boundary value problem associated to the multi-dimensional quantum hydrodynamic model in a bounded domain is
proved. The model consists on conservation of mass equation and
 a momentum balance equation equivalent to  a compressible Euler
equations corrected by a dispersion term of the third order in the
momentum balance. The proof is based on a-priori estimates for the
energy functional for a new observable constructed with an auxiliary
function, and it is shown that, under suitable boundary conditions and assumptions on the initial data, the solution blows up after a finite time.

\end{abstract}
\vspace*{3mm} \noindent \underline{Keywords:}\quad Quantum
hydrodynamic equations, blow up of smooth solutions.
\section{Introduction}
\setcounter{equation}{0}
The evolution of a quantum fluid in a first approximation can be described by a dispersive perturbation associated to the Hamilton-Jacobi system for compressible fluid dynamics, sometimes referred as a dispersive perturbation of the Eikonal equation for the evolution of amplitude and phase velocity of quantum wave guides. This system consists in the continuity equation for the
particle density $\rho$ and for the momentum $\rho u$
\begin{align}
\rho_t+\textrm{div}(\rho u)&=0,\quad \quad\quad t>0,\; x\in\Omega\subseteq \R^d,
\label{conservation_mass_D}\\
(\rho u)_t+ \textrm{div}(\rho u\otimes u)+ \nabla P(\rho)&=\frac{\varepsilon^2}{2}
\rho\nabla\left(\frac{\Delta\sqrt{\rho}}{\sqrt{\rho}}\right),\label{conservation_momentum_D}
\end{align}
subject to the initial conditions
\begin{align}
\rho(\cdot,0)=\rho_I(\cdot)>0,\quad u(\cdot, 0)&=u_I(\cdot),\label{initial_data_QHD}
\end{align}
where $P(\rho)>0 $ describes a pressure function, due to the boundary or to mean field effects, and $\varepsilon$ denotes the scaled
Planck constant.\\
Formally equations (\ref{conservation_mass_D}), (\ref{conservation_momentum_D}) can be
derived through the Madelung's
transform \cite{MD} from the nonlinear Schr\"odinger equation
\begin{align}
i\varepsilon\psi_t=-\frac{\varepsilon^2}2\Delta\psi&+h(|\psi|^2)\psi,\quad x\in\R^d,
\label{schroedinger}\\
\psi(\cdot,0)&=\psi_I(\cdot) \label{initial_data_Schroedinger}
\end{align}
for the wave function $\psi(x,t)$, where $h$ is an integrable function such that
\begin{eqnarray}\label{enthalpy}
h'(\rho)=\frac{P'(\rho)}{\rho}.
\end{eqnarray}
Equation (\ref{schroedinger}) has been proposed as multi-particle
approximations in the mean field theory of quantum mechanics, when
one considers a large number of quantum particles acting in unison
and takes into account only a finite number of particle-particle
interactions. In fact the complex-valued wave function $\psi$ can be rewritten as
$$ \psi=\sqrt{\rho}\exp\left(\frac{iS}{\varepsilon}\right), $$
 where $\rho=|\psi|^2$ and $S$ is some phase function. Then by plugging the above formula
 for the wave function to (\ref{schroedinger}) the
imaginary and real parts of the resulting equation give respectively (\ref{conservation_mass_D})
and (\ref{conservation_momentum_D}) with $u=\nabla S$. The formal equivalence between
(\ref{conservation_mass_D}), (\ref{conservation_momentum_D}) and (\ref{schroedinger})
is explained more in details in the following lemma:
\begin{lemma}\label{Madelung_lemma}
Let $\psi$ be a solution to (\ref{schroedinger}), (\ref{initial_data_Schroedinger})
with initial datum $\psi_I:=\sqrt{\rho_I}\textrm{exp}\left(\frac{iS_I}{\varepsilon}\right)$;
 the functions $\rho:=|\psi|^2$ and
 $u:=\varepsilon\frac{\textrm{Im}(\bar{\psi}\nabla\psi)}{|\psi|^2}$ solve
 (\ref{conservation_mass_D})-(\ref{initial_data_QHD}) with initial data
$\rho_I=|\psi_I|^2$ and $u_I=\nabla S_I$ as long as $\rho>0$.\\
Vice versa let $(\rho,u)$ be a solution to (\ref{conservation_mass_D})-(\ref{initial_data_QHD}),
where $u=\nabla S$ and $\rho>0$ in $\Omega,\; t>0$; the function
$\psi:=\sqrt{\rho}\textrm{exp}\left(\frac{iS}{\varepsilon}\right)$ solves (\ref{schroedinger}).
\end{lemma}

In the recent years problem (\ref{conservation_mass_D}), (\ref{conservation_momentum_D})
has attracted a lot of interest and many papers have been published. In \cite{JL} J\"ungel
and Li proved local-in-time existence
of smooth solutions to (\ref{conservation_mass_D}), (\ref{conservation_momentum_D}) for the
one dimensional case with Dirichlet and Neumann boundary conditions for the particle density
$\rho$. Moreover, if the initial data are close enough to the steady-state solution, the
local-in-time solutions are shown to exist globally in time (so called {\em{small}} solutions).\\
In \cite{GJ}, \cite{GJ02} the
authors studied the positive steady-state solutions to the potential
flow of (\ref{conservation_mass_D}), (\ref{conservation_momentum_D}). In that case the one
dimensional stationary problem can be reformulated into
the following form
\begin{align}\label{1.6}
\rho u=J, \quad
\frac{1}{2}u^2+h(\rho)-\frac{\sqrt{\rho}_{xx}}{\sqrt{\rho}}=K,
\end{align}
with appropriate boundary conditions, where $J$ and $K$ are given
constants (in the classical theory of compressible hydrodynamics, $J$ is the mass flux, $K$ the isoenergetic constant and the function $h$ denotes the enthalpy function). For this Quantum Hydrodynamic case the
authors  proved in \cite{GJ}, \cite{GJ02} that, if $K$ is sufficiently small and $J$ large
enough, system (\ref{1.6}) can not have even a weak solution (supercritical
condition). In addition they showed that one can not recover the
entropic solutions of the compressible Euler limit for the supercritical case
unless the model incorporates diffusion effects. This suggests that
a smooth solution to the corresponding evolution problem (\ref{conservation_mass_D}), (\ref{conservation_momentum_D}), for which the iso-energy functional $K$, defined as 
\begin{eqnarray}\label{K}
K(x,t)=\left(\frac{1}{2}u^2+h(\rho)-\frac{\sqrt{\rho}_{xx}}{\sqrt{\rho}}\right)(x,t),
\end{eqnarray}
is small at the boundary of the domain, will blow up in finite time. This result is proven in Theorem \ref{lemma_primo_risultato} in Section \ref{section1dim} and generalized for a multi-dimensional domain in Theorem \ref{theo_multi_D}. \\
The existence of a one-dimensional stationary viscous quantum
hydrodynamic model is shown in \cite{GuJu04}. This model corresponds to a quantum regularization in the Fokker-Planck
collision operator at the Wigner level \cite{ADM06}, which generates a linear second order viscous term in
(\ref{conservation_mass_D}) and (\ref{conservation_momentum_D}),
after a moment method \cite{Gar94}.
This regularization allows for solutions in case when a ``weakly'' supercritical
condition holds. The existence of solutions in the supercritical regimes
for this model is still an open question.\\
In \cite{Gr03} the author proved non-existence of non-constant traveling waves in the
supersonic regime for the Gross-Pitaevskii equation (\ref{schroedinger}) with
$g'(|\psi|^2)=(1-|\psi|^2)$ in any dimension $d\ge 2$; in particular it is shown
that any quasi-stationary solution in the whole space of the form
$\psi= \psi(x-ct,x_2,...,x_d)$,
with finite energy and velocity $c>\sqrt{2}$, is constant. The proof is based on integral
identities for the variable $\eta := 1-|\psi|^2$ and estimates for the energy functional
$E(\psi):= \int_{\R^d} |\nabla \psi|^2 + (1-|\psi|^2)\;dx$ using the Pohozaev identities.\\
Local in time existence of solutions of (\ref{conservation_mass_D}), (\ref{conservation_momentum_D}) in the whole space has been shown in \cite{LiMa04}.\\
 Our method of showing blow-up for a boundary value problem associated to the system (\ref{conservation_mass_D}), (\ref{conservation_momentum_D}) is in essence inspired in the original argument of R. Glassey
\cite{GL}, where he proved finite time blow up of
smooth solutions to the focusing nonlinear Schr\"odinger equation
(\ref{schroedinger}) with large initial data. The main idea of the
proof in \cite{GL} is the study of the nonnegative quantity
$I(t)=\int_{\R^d}|x|^2|\psi(x,t)|^2\,dx.$ More precisely the author showed
that $I(t)$ will be negative if the smooth solution to
(\ref{schroedinger}) exists long enough for the focusing case. The observable $I(t)$ becomes close to zero due to mass concentration at the point $x=0$.\\
This idea has also been used by Sideris \cite{S85} in the study of
the finite time blowing up of smooth solutions to the compressible
Euler equations and by Xin \cite{X97} in the study of compressible
Navier-Stokes equations.\\
In our work the blow up of the system is caused from generation of vacuum almost everywhere due to boundary constrains. One may expect (or conjecture) this effect is caused by mass concentration at the boundary of the fluid domain. Such study of the type of singularity formation remains an open question and it is not addressed in this manuscript.\\
Since we are dealing with a boundary value problem, the weight function $a(x)$, which
defines the new {\emph{observable}} $I(t)=\int_{\Omega}a(x)\rho(x,t)\;dx$, should be chosen
differently from the Cauchy-problem case. It will turn out that the function $a(x)$ must be a concave function
with zero value on a part of the boundary and its form will depend on the domain and on the boundary conditions of the problem.\\

This paper is organized as follows: in Section \ref{section1dim} we investigate the
finite time blow-up of smooth solutions to the
one-dimensional quantum hydrodynamic equations (we assume for simplicity $\varepsilon^2=2$)
\begin{align}
\rho_t+(\rho u)_x&=0,\quad \quad\quad t>0,\; x\in[0,1],\label{conservation_mass}\\
(\rho u)_t+ (\rho u^2+ P(\rho))_x&=\rho\left(\frac{\sqrt{\rho}_{xx}}{\sqrt{\rho}}\right)_x,
\label{conservation_momentum}
\end{align}
with the initial and boundary conditions
\begin{align}\label{initial_cond}
& \rho(\cdot,0)=\rho_I(\cdot)>0,\quad u(\cdot,0)=u_I(\cdot),\\
& \rho_x(0,t)=\rho_x(1,t)=0,\label{neumann_cond}\\
& \left(u^2+\frac{P(\rho)}{\rho}-\frac{\sqrt{\rho}_{xx}}{\sqrt{\rho}}\right)(0,t)=c_1\le 0,\nonumber\\
& \left(u^2+\frac{P(\rho)}{\rho}-\frac{\sqrt{\rho}_{xx}}{\sqrt{\rho}}\right)(1,t)=c_2\le 0,\label{bcII}
\end{align}
where $c_1$ and $c_2$ are negative constants. Assumption (\ref{bcII}) with $c_1$, $c_2\le 0$ means that there is a nonnegative total momentum flux across the boundary, which implies small values for the iso-energy functional $K(x,t)$ at the boundary for all time.\\
Under Neumann (\ref{neumann_cond}) and Dirichlet boundary conditions for the particle density $\rho$ 
$$
\rho(0,t)=\rho_1,\qquad \rho(1,t)=\rho_2 ,\label{*},\quad \rho_1,\; \rho_2>0,
$$
local in time existence of classical solutions and global existence of small solutions to (\ref{conservation_mass})-(\ref{initial_cond})
is proved in \cite{JL}.\\
We point out that the proof of local existence 
of smooth solutions to (\ref{conservation_mass})-(\ref{bcII}) 
remains at this stage of the work an open problem and will be addressed in future work. We believe however that the same
techniques used in \cite{JL} can be applied for this case and 
existence results will read as follows: if the mass flux $J=\rho u$ at the boundary is small enough and the momentum flux
across the boundary is large enough ($c_1$, $c_2>0$ in (\ref{bcII})), system (\ref{conservation_mass})-(\ref{bcII}) will have global in time existence of solutions. On 
the other hand the authors in \cite{GJ} proved that if the total momentum flux is small enough and the mass flux is large, ($c_1=c_2 <0$) problem (\ref{conservation_mass})-(\ref{bcII}) does not have {\em{stationary}} solutions.  Our conjecture therefore is that the proof of existence of smooth solutions to (\ref{conservation_mass})-(\ref{bcII}) with $c_1$, $c_2$ negative will be possible only locally in time.  
The results shown in this 
paper address the fact that, in the latter case, any {\em{smooth}} solution to (\ref{conservation_mass})-(\ref{bcII}) will blow-up in finite time. However our results do not exclude that weaker solutions of the problem may exist for longer time; our interest is related only on solutions $(\rho, u)$ that deliver strictly positive particle density everywhere in the fluid domain. \\   
Following the same idea
mentioned before for the whole space problem in the Schr\"odinger
approach, we are able to show that in a bounded domain
problem (\ref{conservation_mass})-(\ref{initial_cond}) subject to the boundary conditions (\ref{neumann_cond}), (\ref{bcII}) does not
have global-in-time classical solutions: more precisely it will be
proved that under certain assumptions on the initial
data, the nonnegative quantity $ I(t)=\int_0^1 a(x)\rho(x,t)\;dx$
with $a(x)\ge 0$, for all $x\in [0,1]$, will become strictly
negative after a certain finite time $T^*$, where $T^*$ depends
only on the initial and boundary data of the problem. That implies
that smooth solutions $\rho$ can not exist any longer for $t>T^*$.  In this sense we interpret finite time blow-up of a smooth solution.\\
The proof is based on
two important points: the first one relies on a-priori estimates for the total energy functional
\begin{eqnarray}\label{energy_functional}
E(t)=\int_0^1\left[\frac{1}{2}\rho u^2+g(\rho)+(\sqrt{\rho})_x^2\right](x,t)\;dx,
\end{eqnarray}
consisting on kinetic energy, thermodynamical total enthalpy and quantum energy of the system, where $g(\rho):=\int_0^{\rho}h(r)\;dr$.
The second point is related to some assumptions on the iso-energy functional $K(x,t)$, defined as in (\ref{K}), which is bounded by a nonpositive constant at the boundary of the domain for smooth, local in time positive solutions $\rho$ and $u$. Note that (\ref{K}) becomes the classical Bernoulli's law when the dispersive term is absent. It is our aim to stress that the role of the Bohm potential in $K$ is crucial in order to prove lack of existence of solutions globally in time. In the classical fluiddynamical framework, the third order dispersive term derived from the Bohm potential is absent in the conservation of momentum equation and it is well known that in this case the system has well-posed solutions for any time.  \\
Section \ref{moredim} shows how to extend the same method to the multi-dimensional equations. In the last section we give an alternative configuration for the one-dimensional case where system (\ref{conservation_mass})-(\ref{conservation_momentum}) does not support smooth solution for any time.\\
 We show how the weight function $a(x)$ used in the definition of the observable $I(t)$ depends strongly on the domain $\Omega$. The computations are shown in a general domain and two examples are giving at the end of the paper.

\section{Blow-up in one-dimensional space} \label{section1dim}

In order to present in a simple manner the idea involved, we first consider the problem in a one dimensional domain given by a simple interval. The main results of this section are formulated in the following theorems:

\begin{theorem}\label{lemma_primo_risultato}
Let $\rho\in H^1(0,T,L^2(0,1))\;\cap\; L^2(0,T,H^3(0,1))$, $\rho >0$ in $[0,1]\times [0,T]$, $u\in H^1(0,T,L^2(0,1))\;\cap\;
L^2(0,T,H^2(0,1))$, be a small solution to (\ref{conservation_mass})-(\ref{bcII}) with initial conditions $0<\rho_I\in H^1(0,1)$, $u_I\in L^2(0,1)$. \\
Let $u_I$ and $\rho_I$ be such that
\begin{eqnarray}\label{con_uI}
\int_0^1(1-2x)\rho_I u_I\;dx= M_0<0.
\end{eqnarray}

There exists a positive constant $t_0\le T^* <+\infty$ depending only on the initial and boundary data
such that for $t> T^*$ any classical solution $\rho$ does not exist any longer. In particular $\lim_{t\to T^*} \rho(x,t)=0$ almost everywhere.
\end{theorem}

\begin{remark}
One may interpret in the above theorem that classical solutions $\rho$ lose regularity when $t$ approaches $T^*$. This is our interpretation of finite time blow-up.
\end{remark}\\
Theorem \ref{lemma_primo_risultato} is expanded to higher dimensional sets in the next section, also under some existence conditions on the auxiliary function $a(x)$. \\

In order to prove the above theorem, we need some preliminary results.

\begin{lemma}\label{lemma_energia}
Let $u$, $\rho$ be smooth solutions to (\ref{conservation_mass})-(\ref{bcII})
and $E(t)$ and $K(x,t)$ be defined as in (\ref{energy_functional}) and (\ref{K}) respectively. For every $t>0$ it holds
\begin{eqnarray}
E(t)-E(0)+\int_{0}^{t}(u\rho K)(1,s)-(u\rho K)(0,s)\;ds=0.\label{energy_estimate}
\end{eqnarray}

\end{lemma}
\proof
Taking $\phi=u$ as test function in (\ref{conservation_momentum}) and integrating with respect to $x$ over
$[0,1]$ we get
\begin{eqnarray}\label{week_momentum_conservation}
\int_0^1(u\rho)_tu\;dx=-\int_0^1 \left[u(\rho u^2)_x + u(P(\rho))_x- u\rho
\left(\frac{\sqrt{\rho}_{xx}}{\sqrt{\rho}}\right)_x\right]\;dx.
\end{eqnarray}
We multiply now (\ref{conservation_mass}) by the test function $\phi=h(\rho)-
\frac{\sqrt{\rho}_{xx}}{\sqrt{\rho}}-\frac{1}{2}u^2$:
\begin{eqnarray}\label{week_mass_conservation}
\int_0^1 \rho_t \left(h(\rho)-\frac{\sqrt{\rho}_{xx}}{\sqrt{\rho}}-\frac{1}{2}u^2\right)\;dx\\
=-\int_0^1(u\rho)_x\left(h(\rho)-\frac{\sqrt{\rho}_{xx}}{\sqrt{\rho}}-\frac{1}{2}u^2\right)
\;dx.\nonumber
\end{eqnarray}
Since $\rho_x=0$ on the boundary, it holds
\begin{eqnarray*}
\int_0^1 (u\rho)_t u - \frac{1}{2}\rho_t u^2\;dx = \frac{\partial}{\partial t}\int_0^1
\frac{1}{2}\rho u^2\;dx,\\
\int_0^1 h(\rho)\rho_t\;dx = \frac{\partial}{\partial t}\int_0^1 g(\rho)\;dx,\\
-\int_0^1 \rho_t\frac{\sqrt{\rho}_{xx}}{\sqrt{\rho}}\;dx=\frac{\partial }{\partial t}\int_0^1
(\sqrt{\rho})_{x}^2\;dx.
\end{eqnarray*}
Therefore, after adding up the left- and the right-hand side of
(\ref{week_momentum_conservation})
and (\ref{week_mass_conservation}) it follows
\begin{align*}
\frac{\partial E(t)}{\partial t} &= -\int_0^1 u(\rho u^2)_x\;dx-\int_0^1 u(P(\rho))_x\;dx+ \int_0^1
u\rho \left(\frac{\sqrt{\rho}_{xx}}{\sqrt{\rho}}\right)_x\;dx \\
&\quad-\int_0^1(u\rho)_x h(\rho)\;dx+\int_0^1 (u\rho)_x\frac{\sqrt{\rho}_{xx}}{\sqrt{\rho}}
\;dx+\frac{1}{2}\int_0^1(u\rho)_xu^2\;dx\\
&=:I_1+I_2+I_3+I_4+I_5+I_6.
\end{align*}
Noticing that
\begin{eqnarray*}
I_1+I_6&=&-\frac{1}{2}\int_0^1(\rho u^3)_x\;dx,\\
I_3+I_5&=&\int_0^1\left(\rho u\frac{\sqrt{\rho}_{xx}}{\sqrt{\rho}}\right)_x\;dx,\\
I_2+I_4&=&-\int_0^1 (\rho u \;h(\rho))_x\;dx,
\end{eqnarray*}
the thesis follows.

\hfill$\Box$\\

\begin{lemma}\label{theo_principale}
Let $u$, $\rho$ be smooth solutions to (\ref{conservation_mass})-(\ref{neumann_cond})
and $I(t)$ be a nonnegative quantity defined as
\begin{eqnarray*}
I(t):= \int_0^1 x(1-x)\rho\;dx,\quad\forall\; t\ge 0.
\end{eqnarray*}
It holds
\begin{align*}
\frac{\partial}{\partial t}I(t)&=\int_0^1 (1-2x)\rho_Iu_I\;dx-2\int_0^t\int_0^1
(\rho u^2+P(\rho)+2\sqrt{\rho}_x^2)\;dxds\\
&\quad+\int_0^t\rho\left(u^2+\frac{P(\rho)}{\rho}-\frac{\sqrt{\rho}_{xx}}
{\sqrt{\rho}}\right)(1,s)\;ds\\
&\quad+\int_0^t\rho\left(u^2+\frac{P(\rho)}{\rho}-\frac{\sqrt{\rho}_{xx}}
{\sqrt{\rho}}\right)(0,s)\;ds.
\end{align*}
\end{lemma}
\proof
Integrating (\ref{conservation_momentum}) with respect to $x$ and $t$ over $[0,1]\times [0,t]$ we get
\begin{align}
\int_0^1\rho u\;dx& -\int_0^1\rho_I u_I\;dx+\int_0^t\rho\left(u^2+\frac{P(\rho)}{\rho}\right)(1,s)\;ds\label{1stern}\\
&-\int_0^t\rho\left( u^2+\frac{P(\rho)}{\rho}\right)(0,s)\;ds=\int_0^t(\sqrt{\rho}\sqrt{\rho}_{xx}(1,s)-
\sqrt{\rho}\sqrt{\rho}_{xx}(0,s))\;ds.\nonumber
\end{align}
We multiply now (\ref{conservation_momentum}) by $x$ and integrate over space
and time; taking into account that
\begin{align*}
\int_0^t\int_0^1 x\rho \left(\frac{\sqrt{\rho}_{xx}}{\sqrt{\rho}}\right)_x\;dxds
=\int_0^t\sqrt{\rho}\sqrt{\rho}_{xx}(1,s)\;ds
-\int_0^t\int_0^1 x\rho_x
\frac{\sqrt{\rho}_{xx}}{\sqrt{\rho}}\;dxds\\
-\int_0^t\int_0^1\sqrt{\rho}\sqrt{\rho}_{xx}\;dxds=\int_0^t\sqrt{\rho}\sqrt{\rho}_{xx}(1,s)\;ds+2\int_0^t\int_0^1\sqrt{\rho}_x^2\;dxds,
\end{align*}
it holds that
\begin{align}\label{2sterne}
\int_0^1x\rho u\;dx-\int_0^1x\rho_Iu_I\;dx-\int_0^t\int_0^1(\rho u^2+
P(\rho)+2\sqrt{\rho}_x^2)\;dxds\\
= -\int_0^t\rho \left(u^2+\frac{P(\rho)}{\rho}-
\frac{\sqrt{\rho}_{xx}}{\sqrt{\rho}}\right)(1,s)\;ds.\nonumber
\end{align}
Now we take $\phi=x(1-x)$ as test function for (\ref{conservation_mass}):
\begin{align}
\int_0^1x(1-x)\rho_t\;dx=\int_0^1\rho u\;dx-2\int_0^1 x\rho u\;dx.\label{3sterne}
\end{align}
Substituting (\ref{1stern}) and (\ref{2sterne}) in (\ref{3sterne}) we get
\begin{align*}
\frac{\partial}{\partial t}\int_0^1 x(1-x)\rho\;dx&=\int_0^1 (1-2x)\rho_Iu_I\;dx-2\int_0^t
\int_0^1(\rho u^2+P(\rho)+2\sqrt{\rho}_x^2)\;dxds\\
&\quad+\int_0^t\rho\left(u^2+\frac{P(\rho)}{\rho}-\frac{\sqrt{\rho}_{xx}}
{\sqrt{\rho}}\right)(1,s)\;ds\\
&\quad+\int_0^t\rho\left(u^2+\frac{P(\rho)}{\rho}-\frac{\sqrt{\rho}_{xx}}
{\sqrt{\rho}}\right)(0,s)\;ds,
\end{align*}
and the thesis follows.
\hfill$\Box$\\\\

{\em{Proof of Theorem \ref{lemma_primo_risultato}.}}
Directly from Lemma \ref{theo_principale}, using  (\ref{bcII}) and (\ref{con_uI}), we get
\begin{align*}
I(t)\le I(0) +t\int_0^1 (1-2x)u_I\rho_I\;dx= I(0)+M_0 t,
\end{align*}
where $I(0)=\int_0^1 x(1-x)\rho_I\;dx$. It is easy to see that $I(t)<0$ if
$t>T^*:= -\frac{I(0)}{M_0}$, which implies blow-up for the solution $\rho$ at time $t=T^*$.
\hfill$\Box$\\

\section{Multi-dimensional generalization}\label{moredim}

In this section we extend some aspects of the method developed in the previous
section for one space dimension to any space dimension, provided the boundary value problem has a solvable auxiliary problem in the fluid domain
$\Omega\subset \R^d$, $d\ge 1$.\\
 Our argument consists in constructing  a nonnegative {\sl weight
function} $a(x)$ defined over the set $\bar\Omega$ such that
\begin{align}
\textrm{Hess}(a)\;&\textrm{is negative-semidefinite} \;\textrm{in}\;  \Omega, \label{eqn_a}\\
\Delta a &= -g(x) \le 0 \;\textrm{in}\;  \Omega,\label{eqn_1}\\
a&=0 \;\textrm{on}\;\partial\Omega_D
\subseteq\partial\Omega\label{cond1_a}, \\
\nabla a\cdot \nu &=0
\;\textrm{on}\;\partial\Omega_N=\partial\Omega\setminus\partial\Omega_D\,
,
\label{cond1_b}
\end{align}
for which the new observable
\begin{align*}
I(t)=\int_\Omega a(x)\rho(x,t)\;dx\;
\end{align*}
becomes negative in finite time. The function $g(x)$ must be nonnegative and satisfy the following conditions
\begin{align}\label{cond_g}
\Delta g \le 0 \;\textrm{in}\;  \Omega,\quad \frac{\partial g}{\partial \nu}\ge 0 \;\textrm{on}\; \partial \Omega.
\end{align}
Examples of domains where problem (\ref{eqn_a})- (\ref{cond_g}) can be solved are given in Remark \ref{R2}. \\

In the following theorem we show that the auxiliary weight function
$a(x)$  satisfying \eqref{eqn_a}-(\ref{cond_g}) is good enough to create an {\sl
observable} quantity for which  the $d$-dimensional quantum
hydrodynamic  system
(\ref{conservation_mass_D})-(\ref{initial_data_QHD}) will admit
solutions that blow up in finite time.

Due to the constrains on the existence of the auxiliary weight
function $a(x)$,  the system
(\ref{conservation_mass_D})-(\ref{initial_data_QHD}) under
consideration is subject to the following boundary conditions posed
on the fluid domain $\Omega$:
\begin{align}
\frac{\partial a}{\partial \nu}
\left(\frac{\Delta \sqrt{\rho}}{\sqrt{\rho}}+\frac{|\nabla \sqrt{\rho}|^2}{\rho}-\frac{P(\rho)}{\rho}\right)&-(u\cdot\nabla a)
(u\cdot \nu)=c\le 0 ,\quad\textrm{on}\;\partial \Omega_D,\label{cond_strana}\\
&u\cdot \nu =0,\quad\textrm{on}\;\partial \Omega_N,\label{cond_u2}\\
&\frac{\partial\sqrt{\rho}}{\partial\nu}=0,\quad\textrm{on}\;\partial \Omega,\label{cond_neumann}
\end{align}
where $\partial \Omega = \partial\Omega_D \cup \partial\Omega_N$, $\nu$ is the outward
normal unit vector of $\partial\Omega$ and $c$ a nonpositive constant. The boundary condition (\ref{cond_strana}) can appear to the reader in a first moment not completely correct, since the weight function $a(x)$ appears in its formulation and depends on the domain. We will show in Remark \ref{R2} that for sufficiently smooth domains, the geometry of $a(x)$ is such that condition (\ref{cond_strana}) contains exactly the same information as boundary condition (\ref{bcII}) for the one dimensional problem (which means nonnegative flux across the Dirichlet boundary $\partial \Omega_D$).\\
Existence of solutions to (\ref{conservation_mass_D})-(\ref{initial_data_QHD}) is still an open problem. The difficulty in the existence proof is given by the complexity due the multidimensionality of the problem. However non-existence of stationary solution to (\ref{conservation_mass_D})-(\ref{initial_data_QHD}), (\ref{cond_strana})-(\ref{cond_neumann}) is shown in \cite{GJ}.  
\medskip

\begin{theorem}\label{theo_multi_D}
Let $\rho\in H^1(0,T,L^{2}(\Omega))\;\cap\; L^2(0,T,H^{s+2}(\Omega))$, $\rho >0$ in $\Omega\times [0,T]$, $u\in
H^1(0,T,L^{2}(\Omega))\;\cap\; L^2(0,T,H^{s+1}(\Omega))$, for $s>\frac d2$,
 be a solutions to (\ref{conservation_mass_D})-(\ref{initial_data_QHD}),
 (\ref{cond_strana})-(\ref{cond_neumann}). Let $a(x)\ge 0$ be a nonnegative function satisfying (\ref{eqn_a})-(\ref{cond_g}). If $\rho_I$ and $u_I$ are such that
 $\int_\Omega \rho_Iu_I\cdot \nabla a\;dx<0$, there exists a finite time $t_0\le T^* <+\infty$ depending only on the initial and boundary data such that for $t> T^*$ any classical solution $\rho$ does not exist any longer. In particular $\lim_{t\to T^*} \rho(x,t)=0$ almost everywhere.
\end{theorem}
\begin{remark}
This is the extension of Theorem \ref{lemma_primo_risultato} to higher dimensions.
\end{remark}

\proof
We consider the quantity $I(t)=\int_\Omega a(x)\rho (x,t)\;dx$,
where $a$ satisfies (\ref{eqn_a})-(\ref{cond1_b}), (\ref{cond_g}). Using (\ref{conservation_mass}) and
taking into account that $a=0$ on $\partial\Omega_D$ and  $u\cdot \nu=0$ on $\partial\Omega_N$, the time derivative of $I(t)$ reads as follows
\begin{align}
\frac{\partial I(t)}{\partial t} &=\int_{\Omega} a\rho_t\;dx =-\int_{\Omega}
a\;\textrm{div}(\rho u)\;dx\nonumber\\
& =-\int_{\partial \Omega}a\rho u\cdot \nu\;ds +\int_\Omega \rho u\cdot\nabla a\;dx\nonumber\\
&=\int_\Omega \rho u\cdot\nabla a\;dx,\label{derivata_I}
\end{align}
where $\nu$ is the outward normal unit vector of $\partial\Omega$ at $x$. Let
$\nabla a$ be a test function for (\ref{conservation_momentum}):
\begin{align}
\int_\Omega (\rho u)_t\cdot\nabla a\;dx &+\int_\Omega \nabla a\cdot\textrm{div}
(\rho u\otimes u)\;dx +\int_\Omega \nabla a\cdot \nabla P(\rho)\;dx\nonumber\\
&-\int_\Omega
\rho\nabla\left(\frac{\Delta \sqrt{\rho}}{\sqrt{\rho}}\right)\cdot \nabla a\;dx=0=:I_1+I_2+I_3+I_4.\label{weak_form}
\end{align}
Since  $s>\frac d2$ then $\rho(t,\cdot)\in H^{s+2}(\Omega)\hookrightarrow  C^{2}(\Omega) $
and $u(t,\cdot)\in H^{s+1}(\Omega)\hookrightarrow  C^{1}(\Omega)$ are the classical Sobolev embeddings
with the $L^\infty$-norm in the target space. Integrating by parts it holds
\begin{align*}
I_2&=\int_\Omega \nabla a\cdot \textrm{div}(\rho u\otimes u)\;dx\\
&= \int_\Omega\textrm{div}(\rho u\cdot\nabla a u)\;dx-
\int_\Omega\rho u^T\textrm{Hess}(a)u\;dx\\
&=\int_{\partial\Omega_D}\rho(u\cdot\nabla a)(u\cdot \nu)\;ds
 -\int_\Omega\underbrace{\rho u^T\textrm{Hess}(a)u}_{\le 0}\;dx ,
\end{align*}
taking into account (\ref{cond_u2}). In addition, after integration by parts, we get
\begin{align*}
I_3=\int_\Omega \nabla a\cdot \nabla P(\rho)\;dx=\int_{\partial\Omega} P(\rho)
\frac {\partial a}{\partial \nu}\;ds-\int_\Omega P(\rho)\Delta a\;dx,
\end{align*}
and
\begin{align*}
I_4 =&-\int_\Omega \rho\nabla \left(\frac{\Delta \sqrt{\rho}}{\sqrt{\rho}}\right)
\cdot \nabla a\;dx\\
 =&-\int_{\partial\Omega} \sqrt{\rho}\Delta \sqrt{\rho}\frac{\partial a}{\partial \nu}\;ds +
 2\int_\Omega \Delta \sqrt{\rho}\nabla\sqrt{\rho}\cdot \nabla a\;dx +\int_\Omega \sqrt{\rho}\Delta \sqrt{\rho}\Delta a\;dx\\
=&-\int_{\partial\Omega} \sqrt{\rho}\Delta \sqrt{\rho}\frac{\partial a}{\partial \nu}\;ds
-2 \int_{\Omega} \nabla\sqrt{\rho}\cdot\nabla(\nabla a\cdot \nabla \sqrt{\rho})\;dx
+ \int_\Omega \left(g|\nabla \sqrt{\rho}|^2+\frac{1}{2}\nabla\rho\cdot\nabla g\right)dx\\
=&-\int_{\partial\Omega} \sqrt{\rho}\Delta \sqrt{\rho}\frac{\partial a}{\partial \nu}\;ds -2\int_{\Omega}\nabla\sqrt{\rho}(\textrm{Hess}(\sqrt{\rho})\nabla a + \textrm{Hess}(a)\nabla\sqrt{\rho})\;dx\\
&+\int_\Omega g|\nabla \sqrt{\rho}|^2\;dx+\frac{1}{2}\int_{\partial\Omega}\rho\frac{\partial g}{\partial\nu}\;ds -\frac{1}{2}\int_\Omega\rho\Delta g\;dx \\
=&-\int_{\partial\Omega} \sqrt{\rho}\Delta \sqrt{\rho}\frac{\partial a}{\partial \nu}\;ds -\int_{\Omega}\nabla |\nabla \sqrt{\rho}|^2\cdot \nabla a\;dx - 2\int_\Omega \nabla \sqrt{\rho}^T \textrm{Hess}(a)\nabla\sqrt{\rho}\;dx \\
&+\int_\Omega g|\nabla \sqrt{\rho}|^2\;dx+\frac{1}{2}\int_{\partial\Omega}\rho\frac{\partial g}{\partial\nu}\;ds -\frac{1}{2}\int_\Omega\rho\Delta g\;dx \\
=&-\int_{\partial\Omega} \left(\frac{\partial a}{\partial \nu} \left( \sqrt{\rho}\Delta \sqrt{\rho} + |\nabla \sqrt{\rho}|^2\right)-\frac{1}{2}\rho\frac{\partial g}{\partial\nu}\right)\;ds -2\int_\Omega \underbrace{\nabla \sqrt{\rho}^T \textrm{Hess}(a)\nabla\sqrt{\rho}+\frac{1}{4}\rho\Delta g}_{\le 0}\;dx,
\end{align*}
using (\ref{eqn_1}), (\ref{cond_g}) and (\ref{cond_neumann}). Substituting the above computations in (\ref{weak_form}) we obtain
\begin{align*}
\int_\Omega (\rho u)_t\cdot\nabla a\;dx \le &\int_\Omega\rho u^T\textrm{Hess}(a)u\;dx+
\int_\Omega (P(\rho)+2|\nabla\sqrt{\rho}|^2)\Delta a\;dx\\
&+\int_{\partial\Omega_D}\rho\left[ \frac{\partial a}{\partial \nu}
\left(\frac{\Delta \sqrt{\rho}}{\sqrt{\rho}}+\frac{|\nabla \sqrt{\rho}|^2}{\rho}-\frac{P(\rho)}{\rho}\right)-(u\cdot\nabla a)
(u\cdot \nu)\right]\;ds.\nonumber
\end{align*}
This implies
\begin{align*}
\int_\Omega \rho u\cdot\nabla a\;dx\le\int_\Omega\rho_I u_I\cdot\nabla a\;dx.
\end{align*}
Integration of (\ref{derivata_I}) with respect to time and the above inequality yield
\begin{align*}
I(t)=I(0)+\int_0^t\int_\Omega \rho u\cdot\nabla a\;dxd\tau \le I(0)+t\int_\Omega \rho_I
u_I\cdot\nabla a\;dx.
\end{align*}
If $\rho_I$ and $u_I$ are such that $\int_\Omega \rho_I u_I\cdot\nabla a\;dx \le c<0$,
then for $t\ge -\frac{I(0)}{\int_\Omega \rho_I u_I\cdot\nabla a\;dx}$ it holds
$I(t)\le 0$ and the thesis follows.

\hfill$\Box$

\begin{remark}\label{R2}
We finish  by showing examples of domains and observable
$I(t)$ in which, if (\ref{cond_strana}) holds, any solution to
(\ref{conservation_mass_D})-(\ref{initial_data_QHD}),
(\ref{cond_strana})-(\ref{cond_neumann}) blows up in finite time.\\
The construction of the weight function $a(x)$ satisfying (\ref{eqn_a})-(\ref{cond_g})
depends on the shape (geometry) of the fluid  domain and on the
boundary conditions of the system of equations associated to it. \\
Constraint (\ref{cond_g}) is satisfied in particular when $g\equiv c$, with c a nonnegative constant. In this case a solution to (\ref{eqn_a})-(\ref{cond1_b}) can be given by a quadratic polynomial function or by a solution to the Monge-Ampere equation
\begin{align*}
det(Hess(-a))=1,\quad a=0\;\textrm{on}\;\partial\Omega,
\end{align*}
on a smooth convex domain, such that $Hess(a)$ is a diagonal matrix (see \cite{CKNS}
or \cite{GT} for a general survey in the area).
\end{remark}\\

We give two easy examples of domains.\\

{\em{First case: spherical domain.}} We consider a spherical
domain $\Omega:=\{x\in\R^d|\; |x|\le 1\}$ for problem
(\ref{conservation_mass_D})-(\ref{initial_data_QHD}) with boundary
conditions (\ref{cond_strana})-(\ref{cond_neumann}). In this case the
function $a$ is given by $a(x)=1-|x|^2$,
$x\in\Omega$. It holds $u^T\textrm{Hess}(a)u=-2|u|^2$, $\Delta
a=-2d$ on $\Omega$ and $\frac{\partial a}{\partial\nu}=-2$ on
$\partial\Omega$. Condition (\ref{cond_strana}) reads as follows
\begin{align*}
\int_{\partial\Omega}\rho\left[ \frac{\partial a}{\partial \nu}
\left(\frac{\Delta \sqrt{\rho}}{\sqrt{\rho}}-\frac{P(\rho)}{\rho}\right)-
(u\cdot\nabla a)(u\cdot \nu)\right]\;ds\\
=2\int_{\partial\Omega}\rho\left[(u\cdot x)^2+ \frac{P(\rho)}{\rho}-
\frac{\Delta \sqrt{\rho}}{\sqrt{\rho}}\right]\;ds,\quad x\in\partial\Omega.
\end{align*}
\begin{remark}
The one dimensional case corresponds to this example if the interval considered is $[-1,1]$.
\end{remark}

{\emph{Second case: cylinder-like domain.}} Let $\Omega$ be a bounded cylinder-like
domain $\R^d \supset \Omega := [-1,1]\times \Omega_1 $, where $\Omega_1\subset \R^{d-1}$.
We denote with $z=(x,y)$ for $x\in [-1,1]$ and $y\in\Omega_1$. \\
In this case the boundary $\partial\Omega_D=\Omega_1\times\{x=-1\}\cup \Omega_1\times\{x=1\}$
and $\partial\Omega_N=\partial\Omega_1\times(-1,1)$.\\
We choose $a(x,y)=1-x^2$; it holds $\Delta a=-2$ and $u^T\textrm{Hess}(a)u = -2u_x^2$ on $\Omega$, where $u_x$ denotes
the component of the velocity along the x-direction. In this case
(\ref{cond_strana}) becomes
\begin{align*}
&\int_{\partial\Omega_D}\rho\left[ \frac{\partial a}{\partial \nu}
\left(\frac{\Delta \sqrt{\rho}}{\sqrt{\rho}}-\frac{P(\rho)}{\rho}\right)-(u\cdot\nabla a)
(u\cdot \nu)\right]\;ds\\
=&\int_{\Omega_1\times \{x=-1\}}\rho\left(
{u}_x^2+\frac{P(\rho)}{\rho}-\frac{\Delta\sqrt{\rho}}{\sqrt{\rho}}\right)(0,y,s)\,dy\,ds\\
&+\int_{\Omega_1\times \{x=1\}}\rho\left(
{u}_x^2+\frac{P(\rho)}{\rho}-\frac{\Delta\sqrt{\rho}}{\sqrt{\rho}}\right)(1,y,s)\,dy\,ds,
\end{align*}
taking into account that $u\cdot \nu=0$ and $\nabla a\cdot \nu=0$ on $\partial\Omega_N$.

\begin{remark}
The extension of Theorem \ref{lemma_secondo_risultato} to higher dimensional spaces requires more delicate assumptions in the existence for the initial boundary value problem and corresponding extensions to d-dimensions for (\ref{condizione_aggiunta}). It is expected that the same kind of result will follow for the two cases of fluid domains covered in this section.
\end{remark}

\section{Another configuration for possible blow-up}\label{secIII}

In this section we give another possible blow-up configuration
of solutions to (\ref{conservation_mass})-(\ref{conservation_momentum}) under
different conditions at the boundary. We consider the system in one dimensional space subject to zero Neumann boundary conditions for the particle
density (\ref{neumann_cond}) and Dirichlet-type conditions for the 
velocity   
\begin{align}
u(0,t)=u_0,\quad u(1,t)=u_1,\quad \forall\;t\ge 0. \label{bcIII}
\end{align}
Local in time existence of solutions of such problem (\ref{conservation_mass})-(\ref{conservation_momentum}), (\ref{neumann_cond}), (\ref{bcIII})
 can be shown using the same technique as in \cite{JL}. The problem 
of global in time well-posedness for the system is related to the 
lack of maximum principle for the current density $J=\rho u$. 
Formations of vacuum for the particle density implies no bounds 
for the velocity of the fluid. We believe that smooth solutions 
to (\ref{conservation_mass})-(\ref{conservation_momentum}), (\ref{neumann_cond}), (\ref{bcIII}) with initial data such that $E(0)$ is large enough and $K(0)$ is sufficiently small (where $E$ and $K$ are defined as in (\ref{energy_functional}) and (\ref{K}) respectively), will deliver also in a later time small value for $K(\cdot, t)$ at the boundary of the domain (condition (\ref{condizione_aggiunta})). The rigorous proof of such statement is still an open problem.

\begin{theorem}\label{lemma_secondo_risultato}
Let $\rho\in H^1(0,T,L^2(0,1))\;\cap\; L^2(0,T,H^3(0,1))$, $\rho >0$ in $[0,1]\times [0,T]$, $u\in H^1(0,T,L^2(0,1))\;\cap\;
L^2(0,T,H^2(0,1))$, be a solution to
(\ref{conservation_mass})-(\ref{neumann_cond}), (\ref{bcIII}) where $P(\rho)$ satisfies
(\ref{assumption_pressure_I}) and (\ref{assumption_pressure_II}). Let $0<\rho_I\in H^1(0,1)$, $u_I\in L^2(0,1)$ be such that $E(0)=\int_0^1\frac{1}{2}\rho_I
u_I^2+g(\rho_I)+(\sqrt{\rho_I})_x^2\;dx$ is large enough.
If the quantity $K(x,t)$, defined as in (\ref{K}), at the boundary of the domain is negative and bounded in the following way
\begin{align}\label{condizione_aggiunta}
-M\le K(0,t)\le -\alpha \textrm{max}^2(|u_0|,|u_1|),\quad \;\textrm{for all}\;t\in[0,T],
\end{align}
for some constant $M>0$ and $\alpha >1$, then there exists a positive constant $t_0\le T^* <+\infty$ depending only on the initial and boundary data
such that, for $t> T^*$, any classical solution $\rho$ does not exist any longer. In particular $\lim_{t\to T^*} \rho(x,t)=0$ almost everywhere.
\end{theorem}

For the next theorem we need some assumptions on the pressure function $P(\rho)$ of the system.
More precisely we assume that
\begin{align}\label{assumption_pressure_I}
&\frac{P(\rho)}{g(\rho)}\ge \lambda >0, \quad \forall\; \rho\ge 0,\\
&\frac{P(\rho)}{\rho}-h(\rho)\le 0, \quad \forall\; \rho\ge 0.\label{assumption_pressure_II}
\end{align}
\begin{remark}
Assumptions (\ref{assumption_pressure_I}) and (\ref{assumption_pressure_II}) are satisfied for
example by $P(\rho)=\rho^{\gamma}$ for $\gamma>1$ and by any other sum of power functions. In
this case $g(\rho)=\frac{1}{\gamma-1}\rho^{\gamma}$ and $\frac{P(\rho)}{g(\rho)}\ge \gamma-1>0$.
\end{remark}\\

{\em{Proof of Theorem \ref{lemma_secondo_risultato}.}}
We first show that $K(0,t)=K(1,t)$ for all $t>0$.
This is expected since $K(x,t)$ becomes the ``isoenergetic'' constant if stationary states hold and it is a simple consequence of the conservative character of the underlying Euler-type system. In fact, without loss of generality, we can assume that
\begin{eqnarray*}
u(x,t)=S_x(x,t), \quad S(0,t)=C_0,\; S(1,t)=C_1.
\end{eqnarray*}
This is the classical assumption to writing down the quantum hydrodynamic system from the Madelung transforms applied to the wave function, solution to the corresponding Schr\"odinger equations (see Lemma \ref{Madelung_lemma}).\\
Therefore $\int_0^1u(x,t)\;dx=C_1-C_0$. Dividing now (\ref{conservation_momentum}) by $\rho$ and using (\ref{conservation_mass})
we get
\begin{eqnarray*}
u_t+\frac{1}{2}(u^2)_x+\frac{(P(\rho))_x}{\rho}=\left(\frac{\sqrt{\rho}_{xx}}
{\sqrt{\rho}}\right)_x.
\end{eqnarray*}
Integrating the above equation with respect to $x$ over $[0,1]$, it holds
\begin{eqnarray*}
0=\frac{\partial}{\partial t}\int_0^1u(x,t)\;dx=K(0,t)-K(1,t),
\end{eqnarray*}
which implies $K(0,t)=K(1,t)$.\\
Next, taking into account (\ref{assumption_pressure_II}), the following inequality holds
\begin{eqnarray*}
\rho\left(u^2+\frac{P(\rho)}{\rho}-\frac{\sqrt{\rho}_{xx}}{\sqrt{\rho}}\right)\le \rho
\left(\frac{1}{2}u^2+K(x,t)\right).
\end{eqnarray*}
and therefore from Lemma \ref{theo_principale} we get
\begin{align}\label{stima_principale}
\frac{\partial}{\partial t} I(t) \le \int_0^1(1-2x)\rho_I u_I\;dx
-2\int_0^t\int_0^1(\rho u^2+P(\rho)+2\sqrt{\rho}_x^2)\;dxds\\
\quad +\int_0^t\rho(1,s)\left(K(1,s)+\frac{1}{2}u_1^2\right)\;ds+\int_0^t\rho(0,s)\left(K(0,s)+
\frac{1}{2}u_0^2\right)\;ds,\nonumber
\end{align}
where $I(t)=\int_0^1 x(1-x)\rho \;dx$.
We assume that $\textrm{max}(|u_0|, |u_1|)\neq 0$.\\
Using (\ref{condizione_aggiunta}) on the above inequality we get
\begin{align*}
\frac{\partial}{\partial t}I(t) & \le M_0-2\int_0^t\int_0^1(\rho u^2+P(\rho)+
2\sqrt{\rho}_x^2)\;dxds\\
& \quad -\left(\alpha-\frac{1}{2}\right)\textrm{max}^2(|u_0|,|u_1|)\int_0^t
\rho(0,s)+\rho(1,s)\;ds,
\end{align*}
where $M_0:=\int_0^1(1-2x)\rho_Iu_I\;dx$. \\
If $M_0<0$ the thesis follows as in the
previous theorem.\\
If $M_0\ge 0$ we divide the proof in two parts:\\
{\em{First part:}} If either $\int_0^t\rho(0,s)\;ds\ge \frac{2M_0}{\textrm{max}^2(|u_0|,|u_1|)}$ or
$\int_0^t\rho(1,s)\;ds\ge \frac{2M_0}{\textrm{max}^2(|u_0|,|u_1|)}$ for $t>0$, it holds
\begin{eqnarray*}
I(t)\le I(0)- M_0t(2\alpha -2),
\end{eqnarray*}
where $I(0)=\int_0^1 x(1-x)\rho_I\;dx$, and the theorem is proved.\\
{\em{Second part:}} If both $\int_0^t\rho(0,s)\;ds\le \frac{2M_0}{\textrm{max}^2(|u_0|,|u_1|)}$ and
$\int_0^t\rho(1,s)\;ds\le \frac{2M_0}{\textrm{max}^2(|u_0|,|u_1|)}$, then we need the estimates from Lemma \ref{lemma_energia} and it follows
\begin{align*}
E(t)& =E(0)+\int_0^t K(0,s)(\rho(0,s)u_0-\rho(1,s)u_1)\;ds\\
& \ge E(0)-\textrm{max}(|u_0|,|u_1|)\int_0^t |K(0,s)|(\rho(0,s)+\rho(1,s))\;ds.
\end{align*}
We recall that Lemma \ref{lemma_energia} can be proved also for solutions to (\ref{conservation_mass})-(\ref{conservation_momentum}), (\ref{neumann_cond}), (\ref{bcIII}). Using now (\ref{condizione_aggiunta}) to control $K(0,t)$ we get
\begin{align*}
E(t)\ge E(0)-\frac{4M_0M}{\textrm{max}(|u_0|,|u_1|)}\;.
\end{align*}
Assumption (\ref{assumption_pressure_II}) implies that
\begin{eqnarray*}
-2\int_0^t\int_0^1 \rho u^2+P(\rho)+2(\sqrt{\rho})_x^2\;dxds\le -2\textrm{min}(\lambda,2)
\int_0^t E(s)\;ds\;,
\end{eqnarray*}
and finally from the differential inequality (\ref{stima_principale}), it follows
\begin{eqnarray*}
I(t)\le I(0)+M_0t-\textrm{min}(\lambda,2)t^2
\left(E(0)-\frac{4M_0M}{\textrm{max}(|u_0|,|u_1|)}\right)\;.
\end{eqnarray*}
Therefore if $E(0)\ge \frac{4M_0M}{\textrm{max}(|u_0|,|u_1|)}$ and $t\ge T^*$, where $T^*$
depends only on the initial and boundary conditions, we get $\int_0^1x(1-x)\rho\;dx<0$ for all
$t>T^*$ and the theorem is proved.\\
If $u_0=u_1=0$, then
\begin{align*}
\int_0^1x(1-x)\rho\;dx\le \int_0^1x(1-x)\rho_I\;dx +t\left(M_0-2E(0)\textrm{min}
(\lambda,2)\right),
\end{align*}
and therefore if $E_0$ is large enough such that $E_0\ge \frac{M_0}{2\textrm{min}(\lambda,2)}$
the thesis follows once $t>T^*:=\frac{I_0}{2E_0\textrm{min}(\lambda,2)-M_0}$.

\hfill$\Box$


\section*{Acknowledgment}
The authors would like to thank Daniel Matthes (Department of Mathematics, Mainz University, Germany) and Mary Pugh (Department of Mathematics, University of Toronto, Ontario, Canada) for fruitful discussions.

\end{document}